\documentclass[preprint,authoryear,12pt]{elsarticle}

\usepackage{graphicx}
\usepackage{epsfig}

\usepackage{amssymb}
\usepackage{natbib}
\usepackage{color}
\usepackage{tabularx}
\usepackage{float}
\usepackage{multirow}
\usepackage{lscape}
\usepackage{mathrsfs}
\usepackage{amsmath}
\usepackage{color}
\usepackage{hyperref}
\usepackage[figuresright]{rotating}
\usepackage{booktabs}
\usepackage{threeparttable}
\definecolor{blu}{rgb}{0,0,1}
\hypersetup{
  colorlinks,
  breaklinks,
  linkcolor=blu,
  urlcolor=blu,
  anchorcolor=blu,
  citecolor=blu,
  pdfauthor={Y. Xie, R. Nikutta, L. Hao, A. Li},
  pdfcreator={},
  pdftitle={A Tale of Three Galaxies: A ``Clumpy'' View of
    the Spectroscopically Anomalous Galaxies IRAS~F10398+1455,
    IRAS~F21013-0739 and SDSS~J0808+3948},
  pdfstartpage=1
}

\newcommand       \um           {\,\mu{\rm m}}
\newcommand       \mum        {\,{\rm \mu m}}

\newcommand       \K            {\,{\rm K}}
\newcommand       \simali       {\,{\sim}}

\newcommand{\etal}{\textrm{et al.\ }}
\newcommand{\saga}{$\rm S^3$AGA}
\newcommand{\eg}{\textrm{e.g., }}
\newcommand{\sd}{SDSS~J0808+3948}
\newcommand       \km           {\,{\rm km}}
\newcommand       \s            {\,{\rm s}}
\newcommand       \Mpc          {\,{\rm Mpc}}
\newcommand       \simlt        {\lesssim}

\newcommand{\spitzerirs}{{\em Spitzer}/IRS\ }

\newcommand \aap               {\rm Astron. Astrophys.} 
\newcommand \pasj               {\rm Publ. Astron. Soc. Japan.} 
\newcommand \mnras           {\rm Mon. Not. R. Astron. Soc.}
\newcommand \apj                {\rm Astrophys. J.}
\newcommand \apjl               {\rm Astrophys. J. Lett.}
\newcommand \apjs              {\rm Astrophys. J. Suppl. Ser.}
\newcommand \araa             {\rm Annu. Rev. Astron. Astrophys. J.}
\newcommand \nature          {\rm Nature.}
\newcommand \doi                {\rm doi:\ }
\journal{Planetary and Space Science}
\newcommand \C  {\textsc{Clumpy}}
\newcommand \PAHFIT {\rm PAHFIT}

\begin{document}

\begin{frontmatter}

\title{A Tale of Three Galaxies: 
       A ``Clumpy'' View of 
       the Spectroscopically Anomalous Galaxies
       IRAS~F10398+1455, IRAS~F21013-0739 and 
       SDSS~J0808+3948 
       }

\author[label1,label2,label3]{Yanxia Xie$^{\ast}$}

\address[label1]{Shanghai Astronomical Observatory,
                       Chinese Academy of Sciences,
                       80 Nandan Road, Shanghai 200030, China}
\corref{mycorrespondingauthors.}
\cortext[mycorrespondingauthors.]{Corresponding authors.}
\ead{yanxia.xie@pku.edu.cn}
\address[label2]{Department of Physics and Astronomy, 
                 University of Missouri,
                 Columbia, MO 65211, USA}
\address[label3]{Kavli Institute for Astronomy and Astrophysics, 
                 Peking University, Beijing 100871, China}
\author[label4]{Robert~Nikutta$^{\ast}$}
\ead{robert.nikutta@gmail.com} 
\address[label4]{Instituto de Astrof\'isica,
                 Pontificia Universidad Cat\'olica de Chile,
                 306, Santiago 22, Chile}
\author[label1]{Lei~Hao$^{\ast}$}
\ead{haol@shao.ac.cn}
\author[label2]{Aigen Li$^{\ast}$}
\ead{lia@missouri.edu}

\begin{abstract}
  We investigate the dust properties in three spectroscopically
  anomalous galaxies (IRAS~F10398+1455, IRAS~F21013-0739 and
  SDSS~J0808+3948). Their \textit{Spitzer}/IRS spectra are
  characterized by a steep $\simali$5--8$\mum$ emission continuum,
  strong emission bands from polycyclic aromatic hydrocarbon (PAH)
  molecules, and prominent 10$\mum$ silicate emission. The steep
  $\simali$5--8$\mum$ continuum and strong PAH emission features
  suggest the presence of starbursts, while the silicate emission is
  indicative of significant heating from AGNs. 
The simultaneous detection of these two observational properties 
has rarely been reported on galactic scale.
We employ the PAHFIT software to
estimate their starlight contributions, 
and the \C\ model for the components 
contributed by the AGN tori.
  We find that the \C\ model is generally successful in explaining the
  overall dust infrared emission, although it appears to emit too flat
  at the \hbox{$\simali$5--8$\mum$} continuum to be consistent 
  with that observed in IRAS~F10398+1455 and IRAS~F21013-0739.
  The flat $\simali$5--8$\mum$ continuum calculated from the \C\ model
  could arise from the adopted specific silicate opacity of Ossenkopf
  et al.\ (1992) which exceeds that of the Draine \& Lee (1984)
    ``astronomical silicate'' by a factor up to 2  
     in the $\simali$5--8$\mum$ wavelength range. 
  Future models with a variety of dust species incorporated in the \C\
  radiation transfer regime are needed for a thorough understanding of
  the dust properties of these spectroscopically anomalous galaxies.
\end{abstract}


\end{frontmatter}


\section{Introduction\label{sec:intro}}

%
Dust is a ubiquitous feature of the cosmos. 
It is present in a wide variety of astrophysical
systems, including active galactic nuclei (AGNs) 
and starburst galaxies (see Wang et al.\ 2014).
Amorphous silicates and some form of carbonaceous 
materials (e.g., graphite, amorphous carbon,
and hydrogenated amorphous carbon) are thought 
to be the two dominant cosmic dust species.
Silicate dust reveals its presence in AGNs 
and starbursts through the 9.7 and 18$\mum$
spectral features which respectively arise from 
the Si--O stretching and O--Si--O bending 
modes (e.g., see Henning 2010).

In AGNs, some dust is hot enough to generate a strong, {\it flat}
emission continuum in the $\simali$5--8$\um$ wavelength range 
 when the observed flux $F_\nu$ at frequency $\nu$
is plotted against wavelength $\lambda$,  
making it an effective diagnostic tool to distinguish 
AGN-dominated galaxies from starburst-dominated galaxies 
(e.g., see Laurent \etal 2000, Nardini\etal 2008).  
In addition to silicate dust, polycyclic aromatic
hydrocarbon (PAH) molecules also display rich emission bands in the
mid infrared (IR; Tielens 2008).  
These features are usually very weak or absent
in AGNs because PAHs are believed to have been 
destroyed by the harsh radiation field of AGNs 
(e.g., see Roche et al.\ 1991, Voit 1991,
Siebenmorgen et al.\ 2004; 
but also see Alonso-Herrero \etal 2014).
%
The 3.4$\mum$ C--H stretching absorption feature is 
also detected in AGNs which indicates the presence 
of aliphatic, chain-like hydrocarbon dust in AGNs
(e.g., see Imanishi et al.\ 1997, Mason et al.\ 2004). 

Compared to AGNs, starburst galaxies commonly 
display a low, {\it red} $\simali$5--8$\um$ emission continuum,
strong PAH emission features 
and silicate {\it absorption} features
(\eg Brandl \etal 2006, Hao \etal 2007, Smith \etal 2007).
Moreover, starbursts lack the 3.4$\mum$ absorption feature.
This may be related to the fact that starbursts,
particularly in their nuclear regions, 
are often heavily obscured by dust and ice.
In the Milky Way, the 3.4$\mum$ absorption feature 
is seen in the diffuse interstellar medium (ISM), 
but not in dense clouds where the 3.1$\mum$ H$_2$O
ice absorption feature is strong
(see Pendleton \& Allamandola 2002). 
Starbursts often exhibit strong H$_2$O ice absorption
at 3.1 and 6.0$\mum$ (e.g., see Spoon et al.\ 2004), 
while these ice features are not seen in AGNs
where it is too hot for ice to survive 
against sublimation (see Li 2007).

Very recently, Xie \etal (2014, hereafter paper I) studied the
$\simali$5--40$\mum$ IR spectra of IRAS~F10398+1455,
IRAS~F21013-0739, and SDSS~J0808+3948 obtained with the {\it Infrared
  Spectrograph} (IRS) on board the {\it Spitzer Space Telescope}
(Houck \etal 2004).  They found that the IR spectra of these galaxies
are \textit{anomalous}. Their spectra, on one hand, resemble those of
AGN in the sense that the silicate features in these galaxies are seen
in \textit{emission}; however, they also exhibit strong PAH emission
features which are usually absent in 
AGNs that show similar silicate emission strengths as 
in these three galaxies (e.g., see Spoon \etal 2007).
On the other hand, they are like starbursts in the sense that the
$\simali$5--8$\mum$ emission continua of these galaxies steeply
rise with wavelength $\lambda$, and they show strong PAH emission
features. The silicate features seen in \textit{emission} in these
galaxies are often seen in \textit{absorption} in
starbursts. Furthermore, the \textit{steep} $\simali$5--8$\mum$
emission continuum seen in these galaxies is much \textit{flatter} in
AGNs (see Figure~2).
Let $\alpha = d\ln F_\nu/d\ln\lambda$ be the slope of the
$\simali$5--8$\mum$ emission continuum.
On average, $\alpha \approx 0.8$ for
 quasars as derived from the averaged quasar spectrum 
(see Hao \etal 2007),  
while for IRAS~F10398+1455, IRAS~F21013-0739, and SDSS~J0808+3948,
$\alpha \approx$\,4.1, 4.2, and 4.6, respectively.
A detailed description of the spectral properties and 
a comparison of the {\it Spitzer}/IRS spectra of 
these three galaxies 
with that of typical starbursts and quasars 
can be found in Paper I.

Based on an extensive exploration of the multi-wavelength
observational data, Y.~Xie et al.\ (2016, in preparation) suggest that
young/weak AGNs might be present in these three galaxies.  The focus
of this paper is to examine if the dust distribution in the
circum-nuclear torus, and/or the viewing angle, play a role in
generating such an anomalous mid-IR spectral energy distribution
(SED). This paper is organized as follows.  We describe the data in
\S\ref{sec:data}.  In \S\ref{sec:model} we describe the \C\ torus
model.  The model-fitting to the observed IR emission is presented in
\S\ref{sec:results}.  In \S\ref{sec:discussion} we discuss our
results, and the major conclusions are summarized in
\S\ref{sec:summary}.  Throughout the paper, we assume a cosmological
model with $H_{0} = 70\,h_{70}\km\s^{-1}\Mpc^{-1}$, $\Omega_{m} = 0.3$
and $\Omega_{\land} = 0.7$. $\rm L_{\odot}$ represents solar
luminosity of $\rm 3.826\times10^{33}\,erg\,s^{-1}$.

\section{Observations and Data\label{sec:data}}
 These three galaxies were found in the {\it SDSS-Spitzer
 Spectral Atlas of Galaxies and AGNs} 
(hereafter \saga, Hao et al.\ in preparation).  
We tabulate their basic properties in
Table~\ref{tab:basicpara}.  
The \saga\ sample is constructed by
cross-matching the 7th SDSS Data Release 
of main galaxy sample (Strauss et al.\ 2002) 
and the {\it Cornell AtlaS of Spitzer/IRS Sources} 
(CASSIS) catalog (Lebouteiller et al.\ 2011), 
including 598 galaxies with the SDSS 
and {\it Spitzer}/IRS position-difference smaller than 
$3^{\prime\prime}$.  
In \saga, the mid-IR spectra are restricted 
to those which have a full coverage 
in the {\it Spitzer}/IRS low-resolution 
wavelength interval of $\simali$5--38$\mum$.  
We visually exclude those that have poor quality spectra.  
The redshifts of these galaxies range 
from $z\approx0.015$ to 0.36.  
So far, \saga\ is the largest uniform sample\footnote{%
  By uniform we mean the sample is constructed 
  utilizing only the SDSS and Spitzer/IRS data and excluding the 
  ISO or other IR data. This ensures that the systematic uncertainty 
  of the whole sample is uniform.
  }
which combines optical and mid-IR data at $z \textless 0.4$.
Among this large sample, we have only identified 
three galaxies that exhibit such unique mid-IR 
dust features, implying that they are very rare 
in the local universe.
%
%

\begin{table}[h!]
  {\footnotesize 
  \begin{center} 
  \caption{\footnotesize \label{tab:basicpara}
        Basic parameters for the three spectroscopically anomalous
        galaxies SDSS~J0808+3948, IRAS~F10398+1455, and IRAS~F21013-0739.}
\begin{tabular}{lllccc}
\hline \hline
Sources & R.A. & DEC. & Redshift & Program ID &  $L_{\rm IR}$ \\
              &         &           &               &                     & ($\rm \log_{10}L_{\odot}$) \\
\hline
SDSS~J0808$+$3948  & 08h08m44.27s & $+$39d48m52.36s  & 0.091  & 40444& 10.84 \\ 
IRAS~F10398$+$1455 & 10h42m33.32s & $+$14d39m54.1s  & 0.099  & 40991& 10.50\\
IRAS~F21013$-$0739  & 21h03m58.75s & $-$07d28m02.5s  & 0.136 & 40444& 10.53\\
\hline
\end{tabular}
\end{center}
}
\end{table}

\begin{figure}[h!]
\centering
\vspace{-0.0in}
\includegraphics[angle=0,height=6.in]{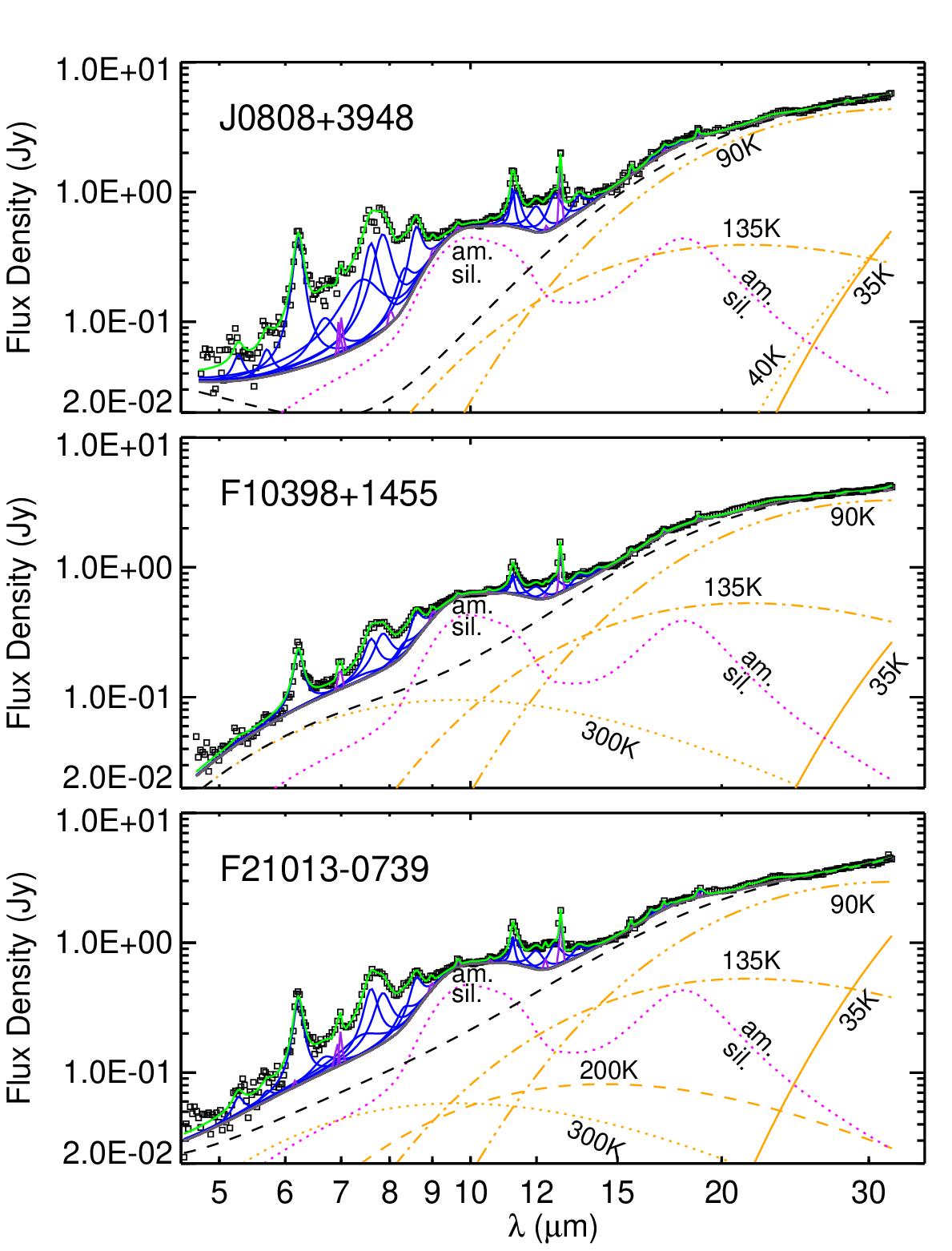}
\vspace{-0.1in}
\caption{\footnotesize
\label{fig:pahfit_3saws}
  Fitting the {\it Spitzer}/IRS spectra of \sd,
  IRAS~F10398+1455, and IRAS~F21013-0739 with \PAHFIT.  The IRS
  spectra (open black squares) are fitted with a combination of PAHs
  (solid blue lines), ionic lines (solid purple lines), 
  warm silicate emission (dotted magenta line), 
  modified blackbodies of different temperatures (orange lines) 
  and starlight.  The sum of the modified
  blackbodies and starlight represents the continuum underneath the
  PAH and silicate features (black dashed line).  In each panel, the
  PAH features are fitted with Drude profiles and the ionic lines are
  fitted with Gaussian functions. 
   }
\end{figure}

\begin{figure}[h!]
\centering
\vspace{-0.0in}
\includegraphics[angle=0,height=6.0in]{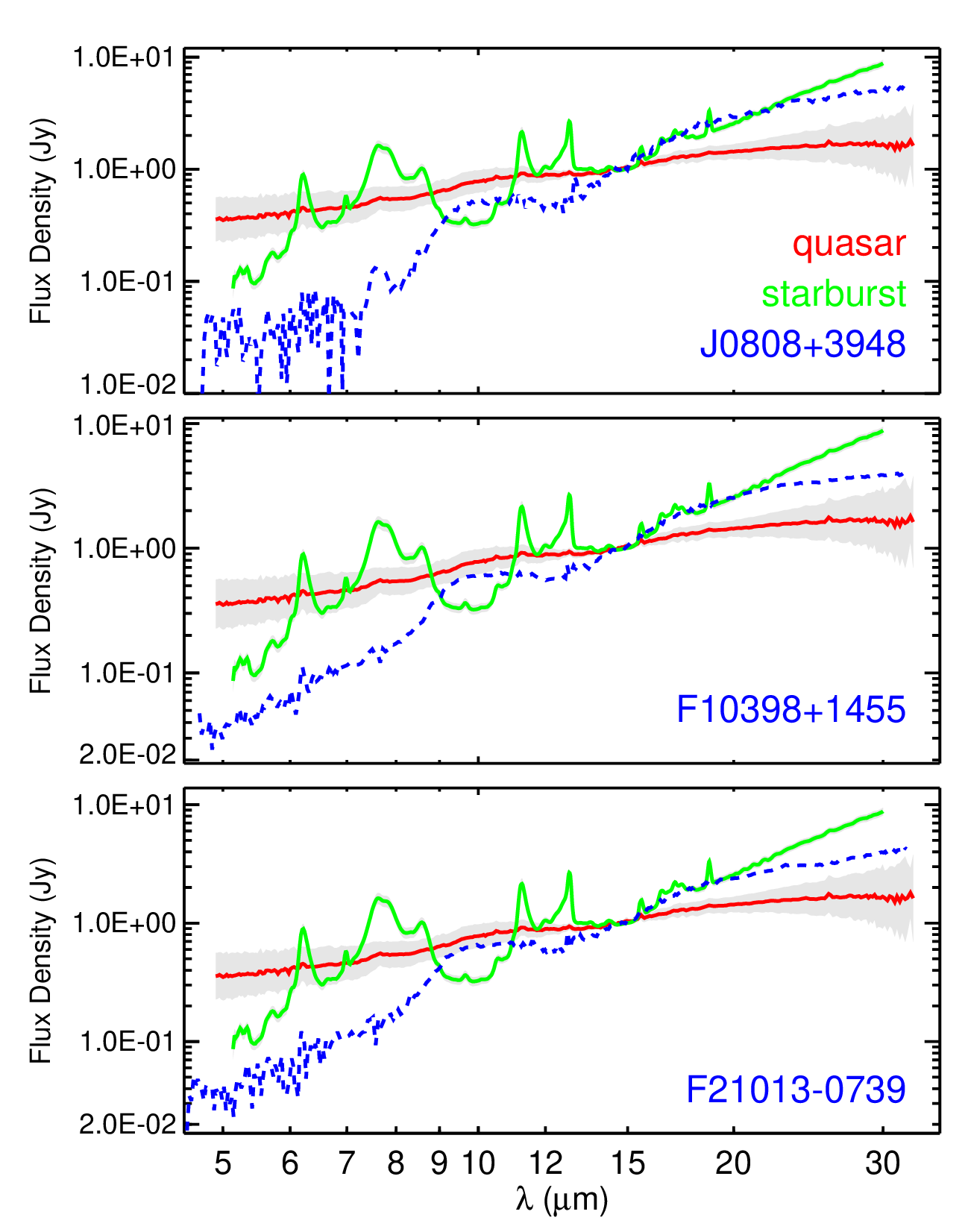}
\vspace{-0.1in}
\caption{\footnotesize
  \label{fig:quasar_sb_3saws}
  The ``residual'' dust emission of \sd, IRAS~F10398+1455, and
  IRAS~F21013-0739 obtained by subtracting from the
  \textit{Spitzer}/IRS spectrum the starlight, PAH and ionic emission lines
  determined via \PAHFIT\ (blue dashed line). In each panel, we also
  show the average IRS spectrum of quasars (red line, Hao \etal 2007),
  and starbursts (green line, Brandl \etal 2006) for comparison.  All
  spectra are normalized at $\lambda = 15\mum$.  
  }
\end{figure}

%
The $\simali$5--8$\mum$ continuum and the 9.7 and 18$\mum$
silicate emission features result from dust grains that are heated
by the AGN central engine. They attain an equilibrium temperature
through the energy balance between the absorption of AGN UV photons
  and the emission of IR photons (see Li 2007).  The PAH emission
  features arise from PAH molecules which are stochastically heated by
  single stellar photons from their host galaxies (see Draine \& Li
  2001).  Due to the differences in their emission mechanisms,
  chemical carriers and emitting regions, and to facilitate a
  detailed examination of the dust emission, we first remove the
  starlight, PAH emission features and the ionic emission lines from
  the \textit{Spitzer}/IRS spectra of these galaxies.  This requires
  an estimation of the dust emission continuum underneath the PAH and
  ionic emission lines.
We approximate the continuum as a sum of starlight and blackbodies of
different temperatures using the \hbox{\PAHFIT}
software (Smith et al.\ 2007) but with the addition of
a warm silicate emission component 
to account for the 10 and 18$\mum$ 
silicate emission features 
(see Paper I for details).
The continuum determined in this way is hereafter referred to as ``the
\PAHFIT\ continuum'', we show our results in
Figure~\ref{fig:pahfit_3saws}.  
We are mostly interested in the residual dust emission obtained by
subtracting the starlight, PAH and ionic emission lines from the observed
\textit{Spitzer}/IRS spectra.
For illustration, we show in Figure~\ref{fig:quasar_sb_3saws} the
residual dust emission of \sd, IRAS~F10398+1455 and IRAS~F21013-0739,
obtained by subtracting
the starlight, PAH and ionic emission lines determined from the \PAHFIT\
approach.

\section{Dust Torus Model\label{sec:model}}
%
  Most of the circum-nuclear AGN dust emission is generated in
  the near-IR, presumably in a geometrically and optically thick torus
  that surrounds the central engine.  The torus exists between
  $r_{\rm in}$ and $r_{\rm out}$, where $r_{\rm in}$ is the dust
  sublimation radius at which dust sublimates or evaporates (which is
  typically a fraction of a parsec and set by the illuminating
  luminosity), and $r_{\rm out}$ is the outer radius which is not well
  constrained.  At $r_{\rm out}$, the torus either gradually fizzles,
  or possibly connects to regions that are dynamically related to the
  host galaxy.  This outer radius is still the matter of
  investigation, both in the light of theoretical differences between
  the smooth-density torus models (e.g., see Pier \& Krolik 1992,
  Granato \& Danese 1994, Efstathiou \& Rowan-Robinson 1995,
  Schartmann \etal 2005, Fritz \etal 2006), and the Clumpy torus
  models (e.g., see Nenkova \etal 2002, 2012a+b, H{\"o}nig \etal 2006,
  Schartmann \etal 2008, Stalevski \etal 2012, Siebenmorgen \etal
  2015), but also because with ALMA it may be at last possible to
  determine $r_{\rm out}$ directly (see Nenkova \etal 2008b, and
  preliminary ALMA observations by Garc{\'{\i}}a-Burillo \etal 2014).

  Very-high angular resolution interferometry of nearby AGN
  suggests that the near- and mid-IR emission is spatially compact
  (e.g. Jaffe \etal 2004, Poncelet \etal 2006, Tristram \etal 2007,
  Raban \etal 2009, Burtscher \etal 2013), albeit this is only derived
  from simple geometric models of the 2D brightness distribution in
  the sky.  The so-modeled mid-IR sizes, a few parsecs, suggest small
  tori indeed, which from the theory/modeling side favors clumpy dust
  distributions.  Acknowledging that the specific choice of preferred
  model is a question beyond the scope of this paper, we decided on
  \C\ SEDs\footnote{\url{www.clumpy.org}} for modeling the near- and
  mid-IR spectra of our three galaxies.  \C, first introduced by
  Nenkova \etal (2002), and fully developed in Nenkova \etal
  (2002a,b), can compute the radiative transfer solution for an
  assumed axially-symmetric distribution of dusty clouds, each opaque
  with optical thickness $\tau_V$, and illuminated by a central UV
  source.  While probably all other models of clumpy tori are
  Monte-Carlo (MC) in nature, \C\ is a model with analytically
  computed statistical properties. It is therefore fast. The downside
  is that it computes the mean solution of all clumpy representations
  of a system at once, and can by design not generate a single clumpy
  view. The single-shot views in MC-based codes, on the other hand,
  can of course be on occasion very far from typical.

  It is important to note that the gas and dust distributions
  in real systems are probably more complex than those in most simple
  1D and 2D models.  More recent 3D MC models allow for additional
  complexity, for instance by introducing a two-phase medium: a clumpy
  component, plus a smooth intra-cloud matter distribution
  (e.g. Stalevski \etal 2012, Siebenmorgen \etal 2015).  Others have
  undertaken great efforts to compute in radiative-hydrodynamical
  simulations the distribution and flow of the interstellar molecular
  gas around a supermassive black hole, down to the very centers of
  AGNs (e.g. Wada \& Norman 2002, Wada \etal 2009, Schartmann \etal
  2010, Hopkins \etal 2012).  

The formalism underlying the \C\ model was in detail described in
Nenkova \etal (2002, 2008a, 2008b). We content ourselves with giving
here the gist. In this model the AGNs torus is populated by dusty
clouds, with matter-free space between them. Radiation can propagate
unimpeded between the individually optically thick clouds. The
visibility of the nucleus along the line-of-sight (LOS) to an external
observer is probabilistic, and, in stark contrast with smooth-density
torus models, is governed not only by the torus angular width and the
viewing angle, but also by the random clumpiness of the cloud
distribution. An observer has a finite chance of seeing the unobscured
nucleus even at edge-on views, if the photons generated in the central
engine manage to avoid absorption by intervening torus clouds. Because
clouds located at different position angles around the AGN show
different phases of their hot and cool faces, an external observer
always registers emission from a mix of different dust temperatures,
originating from co-located regions around the AGN. 
That these different dust temperatures 
co-exist at the same distance from the central engine 
is a hallmark characteristic of clumpiness.  
It has been confirmed observationally in NGC\,1068,
using the {\it Mid-Infrared Interferometric 
Instrument} (MIDI) on the ESO's 
{\it Very Large Telescope Interferometer} (VLTI), 
where two co-located dust components 
at distinctly different temperatures
of $\simali$800\,K and $\sim$300\,K 
have been found (Jaffe \etal 2004, 
Raban \etal 2009, K\"ohler \& Li 2010). 
Notably, the hot component is missing in
the resolved VLTI observations of the Circinus galaxy 
(Tristram \etal 2007, 2014). 
More recently, in a much more complete sample 
of AGNs observed with VLTI/MIDI, 
Burtscher \etal (2013) have modeled
the observed radial brightness profiles 
with a resolved Gaussian 
plus an unresolved point-source. 
They find that the fractional contribution 
of the unresolved component exceeds 
$\simali$50\% for 15 out of 23 detected AGNs. 
Even with the extremely high spatial resolution 
afforded by VLTI/MIDI, 
only two of the 20+ observable AGNs are close enough 
and luminous enough for their very central torus regions 
to be resolved. Burtscher \etal (2013) suspect that this 
may be the reason why so far only NGC\,1068 has revealed 
a very hot dust component.
%
%

  The simplest possible general makeup of an axially symmetric
  clumpy cloud distribution arguably requires no less than six free
  parameters. The radial torus size $r_{\rm out} = Y r_{\rm in}$ is a
  multiple of the scale-invariant dust sublimation radius
  $r_{\rm in}$, which is set only by the AGN luminosity, with $Y$ a
  free parameter.  

The mean number of clouds per radial ray in the equatorial plane
clouds is $N_0$. For radial rays $90-i$~degrees away from the equator
the number of clouds along the line of sight\,(LOS) falls off as a 
Gaussian with width $\sigma$~deg (measured from the equatorial plane). 
Our viewing angle $i$ onto the system is measured from the torus axis. 
The local cloud number density falls off radially as $1/r^q$, 
with the power-law index $q$. Finally, the only property of a single cloud 
(in this minimal model) is its optical depth $\tau_V$ in the visual band.

We consider a mixture of amorphous silicate
(Ossenkopf et al.\ 1992) and graphite (Draine \& Lee 1984),
with a mass mixing ratio of 0.53\,:\,0.47.
The dust grain size is assumed to follow 
a MRN-type power-law distribution for each dust species  
$dn/da \propto a^{-\beta}$,  
where the power-index $\beta$ is fixed at 3.5,
and the lower and upper cut-off sizes are taken to be
$a_{\rm min}\,=\,0.005\,\mum$ and 
$a_{\rm max}\,=\,0.25\,\mum$ respectively
(Mathis et al.\ 1977). 

\section{Results\label{sec:results}} 

In this section we fit the starlight, PAH- and gas-line-subtracted
SEDs of the three galaxies, derived in \S\ref{sec:data}, by using the
\C\ model presented in \S\ref{sec:model}.  We then find the best-fit
model parameters and their marginalized posteriors using the Bayesian
Markov-Chain Monte Carlo (MCMC) SED fitting code developed by Nikutta
\etal (2012).  It performs a multi-dimensional interpolation of the
model SEDs at the requested parameter values and wavelengths, and
allows estimation of model parameters including their significance.
As priors, we assume uniform probability density distributions for all
parameters, in light of their unknown true properties.\footnote{Note
  that we sample the viewing uniformly from $\cos(i) \in [0,1]$ to
  ensure random distribution of orientations in the sphere of the
  sky.} The error bars for F10398+1455 and F21013-0739 are increased
by a factor of 3 and 5 respectively, to aid the fitting algorithm
which we think suffers otherwise under overly optimistic uncertainties
on the data quality.
%

\begin{sidewaystable}[h]
  {\footnotesize 
  \begin{center} 
  \caption{\footnotesize \label{tab:modpara}
               Parameters inferred from fitting with the \C\ models}
\begin{tabular}{lcccccccccccc}
\hline\\[-8pt]
Parameter $\rightarrow$ & \multicolumn{2}{c}{$Y$} & \multicolumn{2}{c}{$\sigma$} & \multicolumn{2}{c}{$q$} & \multicolumn{2}{c}{$\tau_V$} & \multicolumn{2}{c}{$i$} & \multicolumn{2}{c}{$N_0$}  \\
Source $\downarrow$   & MAP & median & MAP & median & MAP & median & MAP & median & MAP & median & MAP & median\\
\hline\\[-8pt]
SDSS~J0808$+$3948  & 20.93 & $22.41_{-1.58}^{+1.59}$ & 29.76   & $30.21_{-2.39}^{+4.96}$  
                                      &   0.00 & $  0.18_{-0.18}^{+0.32}$ & 300.00 & $ 283.09_{-31.42}^{+16.91}$ 
                                      & 90.00 & $  86.18_{-5.18}^{+3.82}$ & 13.80 & $  13.18_{-2.38}^{+1.35}$  \\[5pt] 
IRAS~F10398$+$1455 & 19.40 & $19.42_{-1.75}^{+1.41}$ & 15.00   & $15.34_{-0.34}^{+1.49}$  
                                      &   0.00 & $  0.01_{-0.01}^{+0.09}$ & 184.94 & $185.93_{-11.60}^{+7.74}$ 
                                      & 89.83 & $  89.34_{-2.34}^{+0.66}$ & 10.59 & $   10.82_{-0.49}^{+0.91}$  \\[5pt]
IRAS~F21013$-$0739  & 22.84  & $ 22.60_{-1.77}^{+1.40}$ & 18.58   & $21.03_{-4.20}^{+6.80}$  
                                      &   0.31  & $  0.19_{-0.19}^{+0.31}$  & 300.00 & $274.13_{-32.13}^{+25.87}$ 
                                      & 75.42  & $  76.68_{-4.68}^{+7.32}$ & 15.00  & $10.64_{-3.11}^{+2.96}$ \\[2pt]
\hline
\end{tabular}

\begin{tablenotes}
  Note: For each parameter, the first column indicates the
  maximum-a-posteriori (MAP) value, i.e. the combination of parameter
  values that simultaneously miximizes the likelihood. The second
  column indicates the median value of the parameter distribution
  (i.e. the 50th percentile of the cumulative distribution function,
  CDF), and the $1\sigma$ confidence interval (68.3\% of the CDF
  around the median).
\end{tablenotes}
\end{center}
}
\end{sidewaystable}

\begin{figure}
\centering
\vspace{-0.1in}
\includegraphics[angle=0,height=6.in]{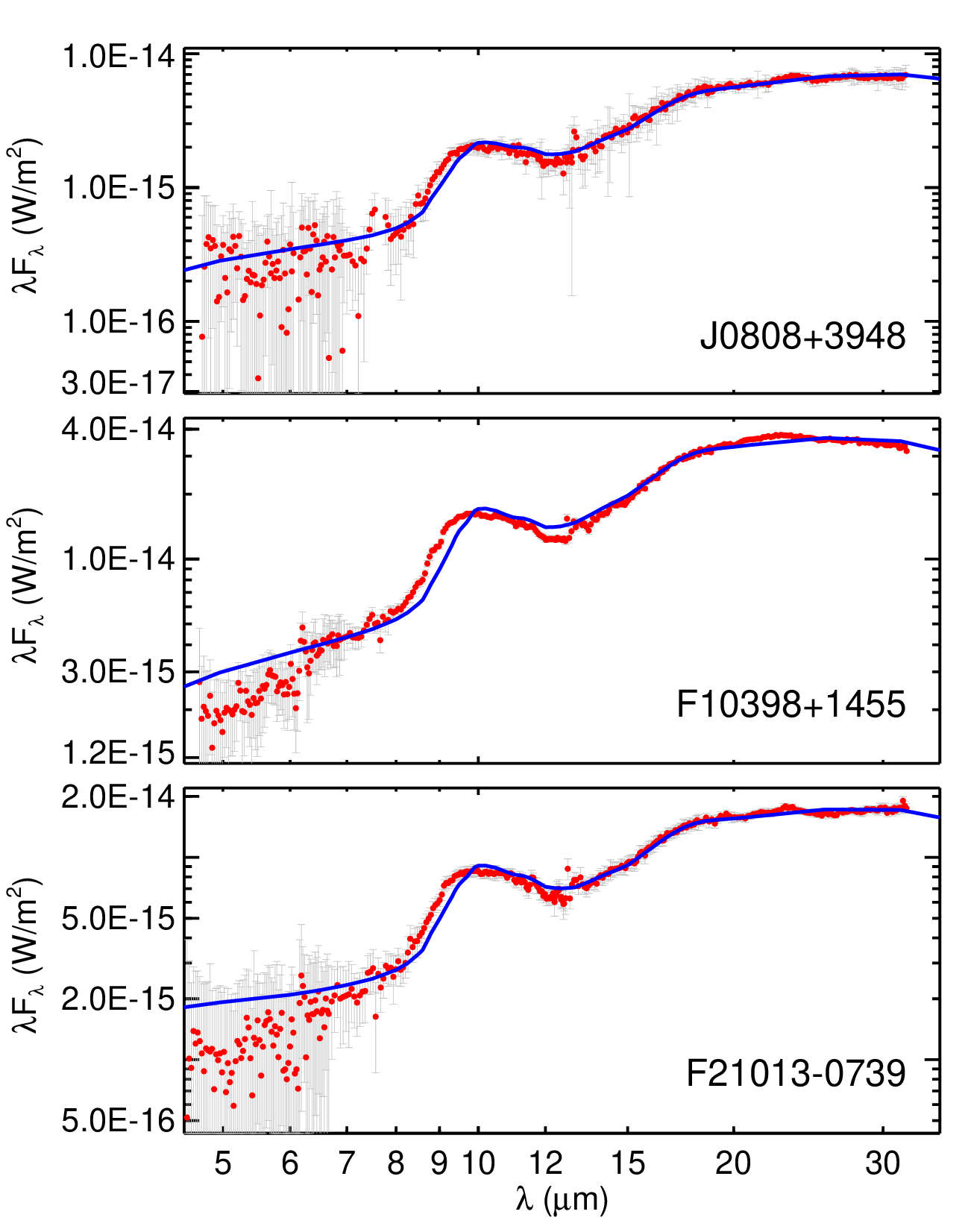}
\caption{\footnotesize
  \label{fig:clumpy}
  \C\ model-fits to the starlight, PAH- and gas-line-subtracted residual spectra
  of \sd, IRAS~F10398+1455 and IRAS~F21013-0739 obtained in
  \S\ref{sec:data}.  In each panel, the red dots plot the observed
  data, and the error bars on the data points are shown with gray
  lines (for F10398+1455 and F21013-0739 they were multiplied by
  factors 3 and 5, respectively. For J0808+3948 they are the 1$\sigma$
  errors). The solid blue line shows the best fit derived from the \C\
  model.}
\end{figure}

We tabulate the best-fit model parameters in
  Table~\ref{tab:modpara} and illustrate the results in
Figure~\ref{fig:clumpy}.  From the best-fitting model it is apparent
that the torus-only model can explain the residual dust SED of \sd\
quite well.  An exception is that the Si--O stretching feature of
silicate dust peaks at $\simali$9.7$\mum$, while the model feature,
owing to the silicate dust opacity used (Ossenkopf \etal 1992), peaks
at $\simali$10.0$\mum$. This is curious, because when {\it Spitzer},
after long expectations, finally detected silicate emission features
in type 1 AGNs (and in fact in some type 2s as well, e.g., see Mason
\etal 2009, Nikutta et al. 2009), they all seemed to be flat-topped
and their peaks shifted appreciably towards longer wavelengths (up to
11.6$\mum$ in some cases).  Researchers have been then calling for
modified dust chemistry in AGN tori (e.g. Sturm \etal 2005) or argued
that radiative transfer effects are the cause of this shift (Nikutta
et al. 2009, but also see Li et al.\ 2008, Smith et al.\ 2010).

For F10398+1455 and F21013-0739, the model predicts too flat a
\hbox{$\simali$5--8$\mum$} continuum to be consistent with
observations. In addition, as in \sd\, the model-calculated silicate
emission feature for F10398+1455 and F21013-0739 also peaks at a
wavelength longer than observed.
These mismatches cannot likely be overcome 
by adjusting the cloud distribution and viewing angle.

%

\section{Discussion\label{sec:discussion}}   
We have shown that no matter how we adjust the torus parameters, the
dust IR emission SED calculated from the \C\ model is too high in the
$\simali$5--8$\mum$ wavelength range compared with the steep continuum
of F10398+1455 and F21013-0739, and that the peak of the model
silicate emission feature is at slightly longer wavelengths for all
three sources.
%

\begin{figure}
\centering
\vspace{-0.1in}
\includegraphics[angle=0,height=6.in]{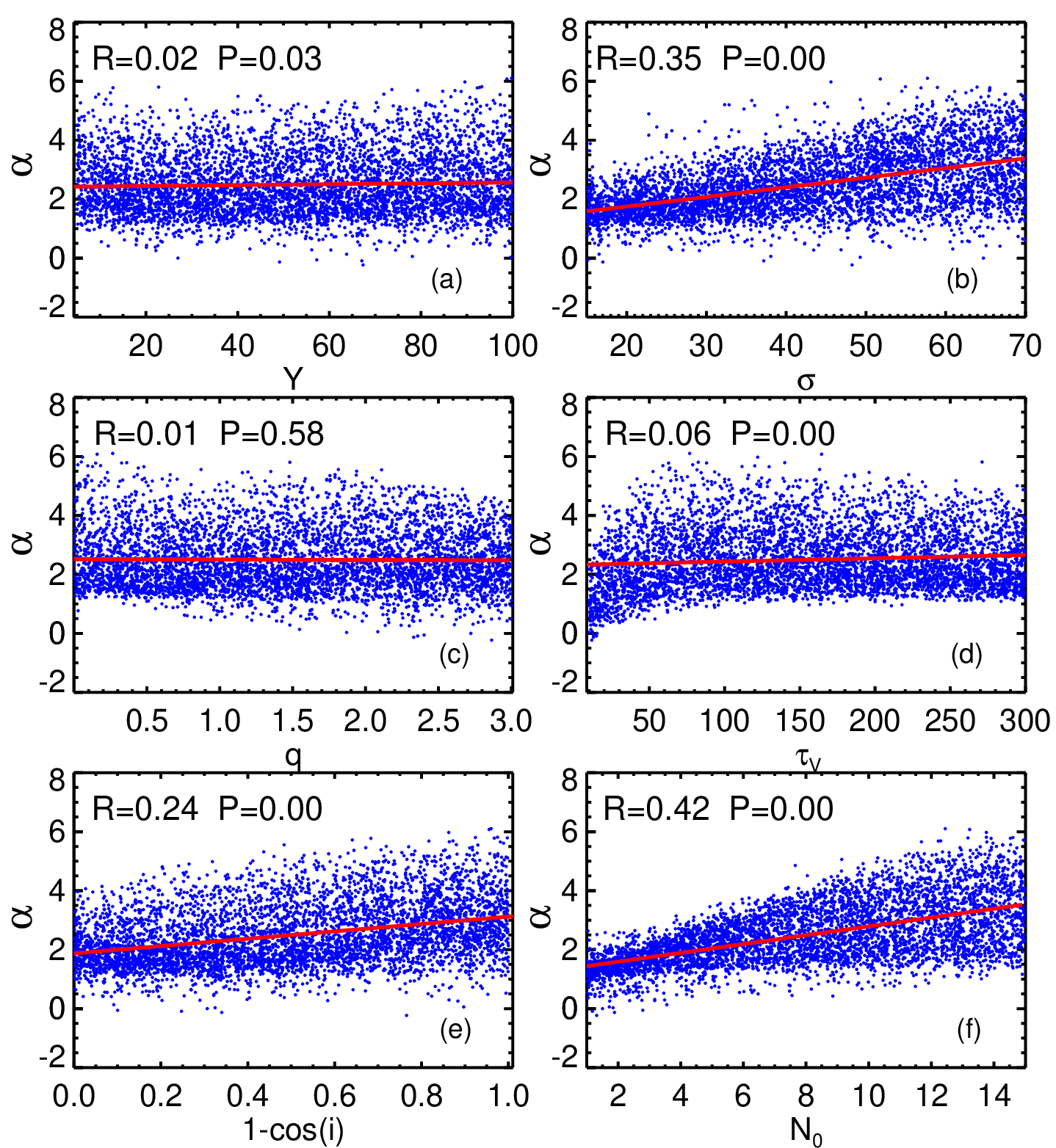}
\caption{\footnotesize
  \label{fig:alpha}
   Correlation of $\alpha = d\ln F_\nu/d\ln\lambda$, 
   the slope of the $\simali$5--8$\mum$ emission continuum,
   with
   (a) $Y=r_{\rm out}/r_{\rm in}$, the ratio of the torus outer radius
       to the torus inner radius;
   (b) $\sigma$, the dispersion of the Gaussian function characterizing 
       the cloud distribution in polar angle (``vertical thickness of
       the torus'')
   (c) $q$, the index of the power-law radial distribution of
       the cloud number density ($\propto 1/r^q$);
   (d) $\tau_V$, the optical depth of a single dust cloud 
       in the visual band;
   (e) $i$, the viewing angle (defined as the angle 
       between the LOS and the torus symmetry axis); and
   (f) $N_0$, the mean number of clouds per radial ray 
       in the equatorial plane.
   While $\alpha$ shows little correlation 
   with $Y$, $q$, $\tau_V$ and $i$,
   it appears that $\alpha$ weakly correlates 
   with $\sigma$ and $N_0$.
    }
\end{figure}


%
One possible reason for the mismatches may lie in the specific
interstellar dust properties that we used: the \C\ model assumes a
mass mixture of 53\% silicate dust and 47\% graphite dust, with the
optical constants of silicate and graphite respectively sourced from
Ossenkopf \etal (1992) and Draine (2003). We note that among those AGNs
exhibiting the silicate emission features, a vast majority have their
``10$\mum$'' feature peaking at a longer wavelength compared to that of
the diffuse ISM (\eg see Hao \etal 2005, Siebenmorgen
\etal 2005, Sturm \etal 2005, Li \etal 2008, Mason \etal 2009, Smith
\etal 2010, Shi \etal 2014). Nikutta \etal (2009) showed that the \C\
model with the Ossenkopf \etal (1992) silicates can explain AGNs with
longward-shifted silicate peaks through radiative transfer effects in
a clumpy medium (which are observed with peaks at up to
11.6$\mum$!). They also make the point that the features in systems with
silicate \emph{absorption} always seem to bottom at 9.7$\mum$. Others
have obtained successful fits using more exotic dust chemistry
(e.g. Jaffe \etal 2004, Sturm \etal 2005).
The Ossenkopf \etal (1992) silicate actually has a longer peak
position of $\simali$10.0$\mum$ compared to that of the interstellar
silicate profile which peaks at $\simali$9.7$\mum$ (Kemper \etal 2004).
However, for these three galaxies of interest in this paper, their
silicate \emph{emission} features all peak at $\simali$9.7$\mum$ and
are comparable to that of the interstellar silicate. Therefore, it is
unavoidable for the \C\ model which is based on the Ossenkopf \etal
(1992) silicate opacity to predict too long a peak wavelength for the silicate
emission feature.
%
Moreover, the $\simali$1--8$\mum$ continuum opacity 
of the Ossenkopf \etal (1992) silicate is higher than 
that of the Draine \& Lee (1984) ``astronomical silicate''
by a factor up to 2. 
This explains the flat $\simali$5--8$\mum$ continuum emission
of the \C\ model: with the 9.7 and 18$\mum$ silicate emission  
features remaining to be fitted, the Ossenkopf \etal (1992) 
silicate opacity would naturally result in a higher level 
of the $\simali$5--8$\mum$ continuum emission than 
the Draine \& Lee (1984) ``astronomical silicate''
(but also see Sirocky et al.\ 2008).
%

Also, the mismatch to the steep $\simali$5--8$\mum$ continuum of
F10398+1455 and F21013-0739 may be caused by the fixed mass ratio of
silicate-to-graphite of $m_{\rm sil}/m_{\rm gra} = 1.13$.  A higher
$m_{\rm sil}/m_{\rm gra}$ ratio may lead to a steep
$\simali$5--8$\mum$ continuum.
In Paper I, we have qualitatively discussed 
the possible dust properties that may account for 
the steep $\simali$5--8$\mum$ continuum 
while exhibiting the prominent 
silicate emission feature 
in terms of dust composition, size and temperature.
Based on the dust opacity characteristics 
presented in Figure~4 of Paper I, 
we proposed that sub-$\mu$m-sized, 
iron-poor silicate as well as 
the deficit of carbon dust could probably
be responsible for the anomalous SEDs
observed in these three galaxies. 

In Xie \etal (2015, hereafter paper II) such dust properties have been
explored in terms of a simple optically-thin model consisting of a
mixture of warm and cold silicate dust, and warm and cold carbon dust.
It was found that models consisting of 
``astronomical'' silicate, amorphous olivine, 
or pyroxene, combined with amorphous carbon 
or graphite, are all capable of successfully 
fitting the observed IR emission.
The dust temperature is 
the primary cause in regulating 
the steep $\simali$5--8$\mum$ continuum 
and silicate emission, 
insensitive to the exact silicate 
or carbon dust mineralogy. 
More specifically, the temperature of 
the $\simali$5--8$\mum$ continuum emitter
(which is essentially carbon dust) of
these galaxies is $\simali$250--400$\K$,
much lower than that of typical quasars
which is $\simali$640$\K$.
The 9.7$\mum$ emission feature constrains
the silicate dust size to not exceed 
$\simali$1.0$\mum$ (i.e., $a\simlt1.0\mum$).
The mass ratio of the warm carbon dust to
the warm silicate dust 
ranges from $\simali$0.2 to $\simali$2.0,
with a mean ratio of $\simali$0.98.
Similarly, 
the mass ratio of the cold carbon dust to
the cold silicate dust 
ranges from $\simali$0.34 to $\simali$2.0,
with a mean ratio of $\simali$1.5.
The total dust mass is dominated 
by the cold components, 
with the warm components only accounting for
$<$\,0.15\% of the total dust mass.

To more effectively explore the dust properties of these three
galaxies, one needs to incorporate these different dust compositions
and mass ratios into the \C\ radiation transfer model.  In Nikutta
\etal (2016, in preparation), we will investigate these dust
characteristics with the \C\ model.  Alternatively, the
  $\simali$5--8$\mum$ continuum observed in the three galaxies might
  be affected by an optically thick dust disk (distinct from the
  clumpy component), which displays steeper $\simali$5--8$\mum$
  continuum emission in comparison with the clumpy models (see
  Siebenmorgen \etal 2015, Figure 5c).

%
To further explore if (and how) the $\simali$5--8$\mum$
  continuum emission is connected with the nature and distribution of
  the dust in a torus, we investigate the correlation between the
  slope of $\simali$5--8$\mum$ continuum (defined as $\alpha$, see
  \S1) and the \C\ model parameters.  We show our results in
  Figure~\ref{fig:alpha}; in each panel we also show the Kendall rank
  correlation coefficient\,(R) and significance\,(P) of correlation as
  well as the linear fit to the data set.  It is apparent that
  $\alpha$ shows little correlation with $Y$, $q$, $\tau_V$ and $i$,
while it appears that $\alpha$ is related to $\sigma$ and $N_0$.
The weak correlation between $\alpha$ and $\sigma$ or $N_0$
is due to the fact that a {\it larger} $\sigma$ or $N_0$
results in {\it more} obscuration in the inner torus
and therefore overall {\it colder} dust emission.    
%

\section{Summary\label{sec:summary}}
We have modeled the \textit{Spitzer}/IRS spectra of three
spectroscopically anomalous galaxies (IRAS~F10398+1455,
IRAS~F21013-0739 and SDSS~J0808+3948) by decomposing them via
\PAHFIT, subtracting the starlight, PAH and atom-line components, 
and then modeling the residual dust component with the \C\ torus model.
The IR spectral characteristics of these galaxies are unique in the
sense that they show silicate emission which is characteristic of AGNs,
while they also show a steep $\simali$5--8$\um$ continuum and strong
PAH emission features typical for starburst galaxies.
In contrast, AGNs exhibit a {\it flat} emission 
continuum at $\simali$5--8$\mum$ and lack the PAH 
emission features. The 9.7 and 18$\mum$ silicate 
features seen in emission in these three galaxies
are seen in absorption in starbursts.
%

We have investigated whether their anomalous SEDs 
could be explained by adjusting the dust cloud distribution 
and the viewing angle toward the AGNs torus. 
We find that the \C\ model is generally successful in explaining the
overall dust IR emission, although it appears to emit too flat a
$\simali$5--8$\mum$ continuum to be consistent with that observed in
IRAS~F10398+1455 and IRAS~F21013-0739.
Also, the peak wavelength of 
the model silicate emission feature 
is too long compared with that observed 
in all three sources.
We argue that the problem of the model-predicted longward-shifted
silicate emission peak could be solved by incorporating different
silicate species other than the Ossenkopf \etal (1992) silicate
opacity, and the problem of the model-predicted flat
$\simali$5--8$\mum$ continuum could be solved by incorporating
iron-poor silicate dust and/or a higher silicate-to-graphite mass
ratio than that currently adopted in the \C\ model. 
Other than modifying the dust minerology, 
the presence of an optically thick dusty disk 
might also produce a steep slope of 
$\simali$5--8$\mum$ continuum 
as seen in the three galaxies. 

\section*{Acknowledgements}
\label{acknowledgements}
We thank all three referees for their very helpful comments.  
LH and XYX are partially supported by the 973 Program 
of China (2013CB834905, 2009CB824800), 
the Strategic Priority Research Program ``The Emergence
of Cosmological Structures'' of Chinese Academy of Sciences
(XDB09000000), the Shanghai Pujiang Talents Program (10pj1411800) and
NSFC 11073040, 11173019.  AL and XYX are supported in part by NSF
AST-1311804 and NASA NNX14AF68G.  RN acknowledges support by FONDECYT
grant No. 3140436.  The Cornell Atlas of \spitzerirs Sources (CASSIS)
is a product of the Infrared Science Center at Cornell University,
supported by NASA and JPL.


\end{document}